\documentclass[a4paper,12pt]{article}
\usepackage{graphicx}
\usepackage{natbib}
\DeclareGraphicsExtensions{.pdf, .eps}
\usepackage{epstopdf}

\begin{document}
\pagenumbering{arabic}

\author{
    Ji\v{r}\'{\i} Hor\'{a}k \& Michal Bursa \\[.5em]
    {\small\em Astronomical Institute, Academy of Sciences, Bo\v{c}n\'{\i} II 1401/1a, 141 31 Praha 4, Czechia}
}
\title{Polarization from the oscillating magnetized accretion torus}
\date{Proceedings of the conference ``The coming of age of X-ray polarimetry'',\\ Rome, Italy, 27--30 April 2009}
\maketitle

\abstract{We study oscillations of accretion torus with azimuthal magnetic field. For several lowest-order modes we calculate eigenfrequencies and eigenfunctions and calculate corresponding intensity and polarization light curves using advanced ray-tracing methods.}

\vspace{4em}

\section{Introduction and model description}

In addition to spectroscopy, polarimetry provides us with further information about a source of radiation and namely about its geometry. In particular, a time-resolved polarimetry may be a useful tool in identifying various types oscillations in non-stationary sources. In this note we explore this idea for a system consisting of a rotating black hole surrounded by a thick accretion disk (accretion torus).
 
\begin{figure}[t]
\centering
\includegraphics[width=0.8\textwidth]{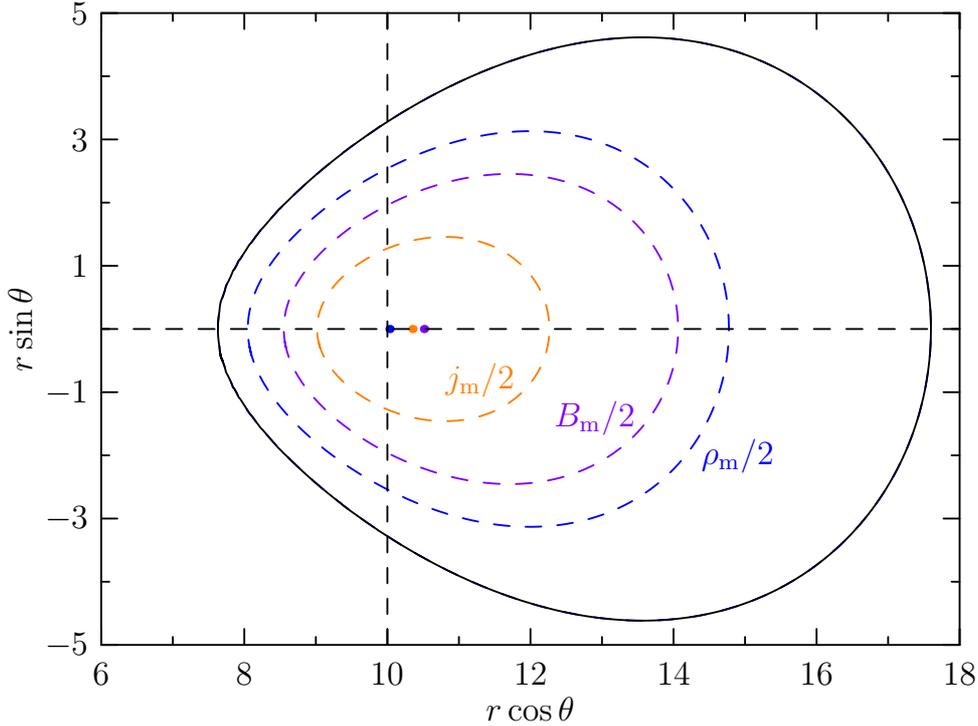}
\caption{The view of the torus shape at its cross-section. The plot shows points, where there is a maximum of density ($\rho_{\rm m}$), magnetic field induction ($B_{\rm m}$) and emissivity ($j_{\rm m}$). Dashed contours illustrate, where the respective quantities gain one half of their maximal values. The solid black line represents the outer boundary of the torus body.}
\label{fig:torus-shape}
\end{figure}

An analytical description of this set-up was first given by \citet{Okada+1989} in Newtonian gravity and recently by \citet{Komissarov2006} in Kerr geometry \citep[see also][]{Montero+2007}. We differ from the latter work only by employing a polytropic equation of state in the form $P = K\rho^{1+1/n}$, but we use the same relation between magnetic pressure $p_{\rm m}$ and the enthalpy $w$, $p_{\rm m} = K_{\rm m} \mathcal{L}^{\mu-1} w^\mu$ with $\mathcal{L}=g_{t\phi}^2 - g_{tt}g_{\phi\phi}$ and $\mu$ being a parameter (\mbox{$\mu\!=\!2$} in our work). We assume that the angular momentum is constant over the whole volume of the torus. The spatial profiles of the fluid density, pressure and the local magnetic field strength measured in a local comoving reference frame can be expressed using Lane-Emden functions $f(r,\theta)$ as (Figure~\ref{fig:torus-shape})
$$
  \rho = \rho_0 f^n,
  \quad
  p = p_0 f^{n+1}
  \quad
  B = B_0 \left(\frac{\mathcal{L}}{\mathcal{L}_0}\right)^{\mu-1}
  \left(\frac{1 + n c_{\mathrm{s}0}^2 f}{1 + n c_{\mathrm{s}0}^2}\right)^\mu f^{\mu n}\;,
$$
where $c_{\mathrm{s}}$ is speed of sound and all quantities with subscript zero are evaluated at the torus center $[r_0,\pi/2]$. At a given point the Lane-Emden function is a solution of the equation
$$
  \mathcal{U}-\mathcal{U}_\mathrm{in} + \ln\left(1 + n c_{\mathrm{s}0}^2f\right) + 
  \frac{\mu n c_{\mathrm{s}0}^2 [1 + n c_{\mathrm{s}0}^2 f]^{\mu-1}}
  {(\mu-1)(n+1)\beta_\mathrm{p}[1 + n xc_{\mathrm{s}0}^2 f]^\mu}
  \left(\frac{\mathcal{L}}{\mathcal{L}_0}\right)^{\mu-1} = 0.
$$
where $\mathcal{U}=\ln u_t$ is the relativistic effective potential, $\mathcal{U}_\mathrm{in}$ is its value at the inner edge of the torus and $\beta_p=p_0/p_{\mathrm{m}0}$ is the plasma beta-parameter (\mbox{$\beta\!=\!1$} in our work).

\section{Polarization from the stationary torus}

\begin{figure}[t]
\centering
\includegraphics[width=0.8\textwidth]{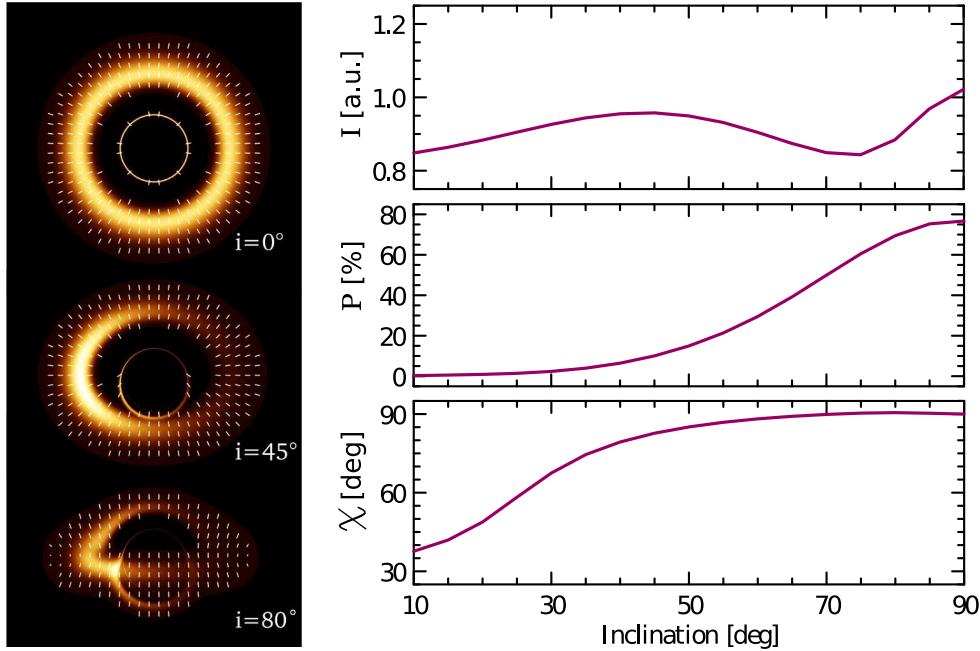}
\caption{The dependence of the total luminosity and of the degree and the angle of polarization on the location of the observer (right). The left part of the figure shows three views of the magnetized torus for inclinations $0^\circ$, $45^\circ$ and $80^\circ$. On top of the color map of the emission intensity, there are line segments representing the polarization properties of radiation at a given place: length of lines means the degree, and orientation indicates the angle of polarization.}
\label{fig:torus-stat}
\end{figure}

Because the fluid is highly magnetized, the dominating emission process in our model is synchrotron radiation that produces polarized light. In the local comoving reference frame, the direction of polarization is perpendicular to the projection of magnetic field onto the polarization plane (i.e.\  the plane perpendicular to the direction of emitted photons). Therefore, in a polarization basis $\{\mathbf{X}, \mathbf{Y}\}$, where the $\mathbf{Y}$ vector makes an angle $\varphi$ with that projection, the frequency-integrated local Stokes emissivities are given by
$$
  J_I = J_0 \left(\frac{\rho}{\rho_0}\right) \left(\frac{B_\perp}{B_0}\right)^{1+\alpha}, \quad
  J_Q = p J_I \cos(2\phi), \quad
  J_U = -p J_I \sin(2\phi),
$$
where $J_0$ is a constant, $B_\perp$ is a normalized projection of the magnetic field onto the polarization plane basis and $p=(\alpha+1)/(\alpha+5/3)$ for the case of the power-law distribution of radiating electrons with spectral index $\alpha$.

Due to the global orientation of the magnetic field, the synchrotron emission from the torus shows certain degree of polarization which depends on the orientation of the torus with respect to a distant observer. When the observer sees the torus from above (along its symmetry axis) the total degree of polarization is zero (or negligible) due to axial symmetry. With increasing inclination the net degree of polarization increases and is the largest for observers looking edge-on (Figure~\ref{fig:torus-stat}).

\section{Polarization changes due to torus oscillations}

\begin{figure}[p]
\centering
\includegraphics[width=0.8\textwidth]{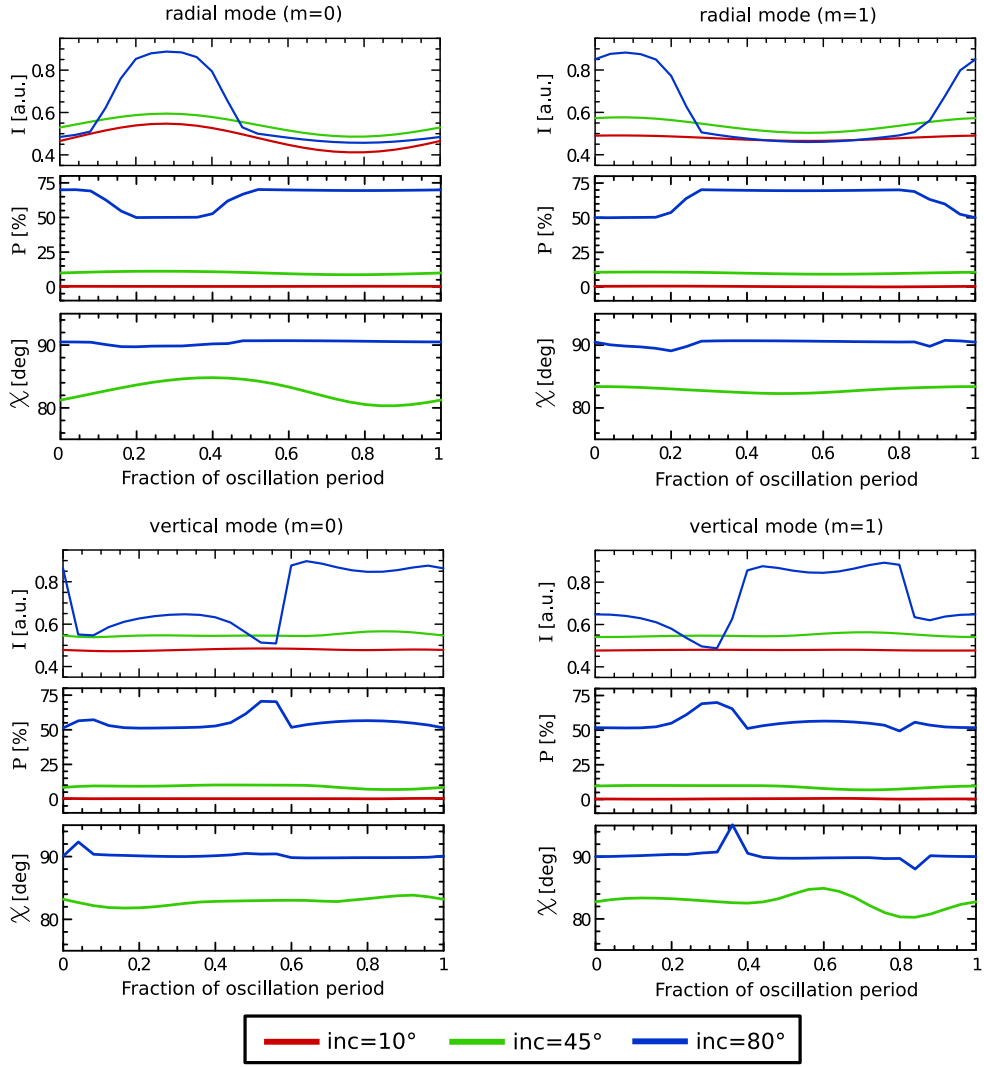}
\caption{The dependence of the total luminosity and of the degree and the angle of polarization on the time phase of some particular oscillation modes: radial and vertical epicyclic modes with wavenumbers $m\!=\!0$ and $m\!=\!1$. Lightcurves for three inclination are drawn: $10^\circ$, $45^\circ$ and $80^\circ$. Horizontal axes show relative phase of each oscillation -- their absolute periods differ. The degree of polarization for $10^\circ$ inclination is close to zero, hence the corresponding polarization angle curve is not plotted.}
\label{fig:torus-pert}
\end{figure}

Accretion tori suffer from various perturbations induced by non-steady accretion of mass. Numerical simulations show that perturbations lead to a development of variety of oscillating modes in the torus body. In the context of polarization it is interesting to examine what effect these internal oscillation have on produced synchrotron radiation, because tracing temporal changes in polarization can be a possible way to probe the geometry of accretion flows.

The equations governing linear perturbations of the torus can solved analytically in the limit of slender torus, when $c_{\mathrm{s}0}\rightarrow0$. In particular, the torus admits two major global modes, radial and vertical epicyclic oscillations, whose poloidal velocity fields are nearly uniform on the torus cross-section, and whose eigenfunctions (in terms of Eulerian velocity perturbation) can be expressed as
\begin{equation}
  \delta u^\alpha = \mathcal{A_\mathrm{r}}\exp[-i(\omega t - m\phi)]\,\delta^\alpha_r,\quad
  \delta u^\alpha = \mathcal{A_\mathrm{v}}\exp[-i(\omega t - m\phi)]\,\delta^\alpha_\theta,
\end{equation}
where $m$ is the azimuthal wavenumber. Frequencies of these oscillations are given by $\omega = \omega_{r,\theta} + m\Omega_0$, where $\Omega_0$ and $\omega_{r,\theta}$ is the angular frequency and radial and vertical epicyclic frequencies at the center of the torus, respectively. Recently, the radial epicyclic mode has been found also in numerical simulations by \citet{Montero+2007}.

Using a numerical ray-tracing technique, we calculate the transport of local Stokes parameters and construct torus light curves as it would appear to a distant observer. We select four distinct oscillation modes, for which we compute time profiles of total intensity, polarization degree and angle. The four modes are radial and vertical epicyclic modes with wavenumbers $m\!=\!0$ and $m\!=\!1$. Resulting lightcurves are summarized in Figure~\ref{fig:torus-pert}. Most dramatic changes happen to lightcurves when the source is seen under large inclinations (edge-on). That is because Doppler boosting, gravitational light bending and other relativistic effects are most pronounced in side view. We find that there is a visible difference in temporal evolution of polarization lightcurves between radial and vertical modes. There are, however, little differences between $m\!=\!0$ and $m\!=\!1$ modes (beside some phase shift) of both radial and vertical oscillations respectively. Still these modes can be distinguished by their different oscillation frequencies.

\section{Conclusions}
In our example, one can discriminate between radial and/or vertical oscillation modes easily. On the other hand differences between axisymmetric and non-axisymmetric modes are less apparent -- the two cases mainly differ in phases but their phase profiles are quite similar.

\subsection*{Acknowledgements}
The authors have got support from Czech MSMT grant LC06014 and from project AV0Z10030501.

\end{document}